\documentclass[12pt]{article}

\usepackage{graphicx}

\usepackage{amssymb,amsmath,amsfonts,amsthm,graphicx,psfrag}
\usepackage{hyperref}
\usepackage{slashed,braket}
\usepackage{color}

\newcommand{\bbraket}[1]{\braket{\braket{#1}}}
\newcommand{\tr}{\text{tr}}

\newcommand{\beq}{\begin{eqnarray}}
\newcommand{\eeq}{\end{eqnarray}}
\newcommand{\be}{\begin{equation}}
\newcommand{\ee}{\end{equation}}

\newcommand{{\cD}}{\mathcal{D}}
\newcommand{{\const}}{\text{~const~}}
\newcommand{{\MeV}}{\text{~MeV~}}
\newcommand{\vep}{\mbox{\boldmath${\rm p}$}}
\newcommand{\lan}{\langle}
\newcommand{\ran}{\rangle}

\begin{document}

\title{Quark condensate  in QCD at nonzero magnetic field and temperature}

\author{  Yu.A.Simonov \\
NRC ``Kurchatov Institute" - ITEP \\
Moscow, 117218 Russia}

\maketitle
\begin{abstract}
The basic form of the quark condensate for arbitrary values of external magnetic field and temperature, is derived using the field equations with account of confinement in the Field Correlator Method.
The resulting expression of the chiral condensate is shown to be proportional to square of the singlet $q\bar q$ ground state wave function at origin, $|\phi_0(0)|^2$. For light quarks without magnetic field the condensate is proportional to $\sigma^{3/2}$, where $\sigma$ is the string tension. Numerical results are presented in 5 Tables and shown to be in good agreement with the lattice data, both for nonzero magnetic field $eB$ and temperature $T$ in the range $0 <T < 120$~MeV,~ $0 <eB <4$~GeV$^2$.
\end{abstract}

\section{Introduction}

The chiral symmetry breaking (CSB) is the fundamental property of the QCD dynamics \cite{1,2}.
The most popular formalism for the study of this phenomenon and the theoretical predictions is
 the chiral perturbation theory (CPTh) \cite{3} using chiral effective Lagrangian .
The primary parameter of CSB is the quark condensate, $ \Sigma_i = | \lan \bar q_i q_i\ran |$,
 which is nonzero in the chiral broken phase. Therefore the most important problem is to find the origin of this parameter and its connection and role in the overall dynamics of QCD,
in particular, the connection of $\Sigma_i$ and the string tension $\sigma$, which provides confinement. The chiral condensate  was intensively studied in the framework of the QCD lattice theory \cite{4,5,6,7,8,9,10} at zero values of temperature and magnetic field. On the theoretical side the main eff0rts in this area have been devoted to the standard chiral theory including  the chiral perturbation theory (CPTh) \cite{3}.
The CPTh in its basic steps does not contain confinement and quark-antiquark degrees
of freedom and is  quite successful in the case of zero temperature and magnetic field. However
this defect strongly limits its possible application to the realistic physical systems
e.g. for chiral mesons or chiral condensate in the external electromagnetic fields where CPTh provides results in strong disagreement with the lattice data \cite{10*}.
This topic has become popular in 2011-2012 when several lattice studies discovered interesting picture of the chiral physics in the external magnetic field and also at nonzero temperature $T$ \cite{11,12,13,14,15,16,17}. The phenomena of increasing chiral condensate in magnetic field (m.f.) were denoted as "the magnetic catalysis" \cite{17*} .

 As an alternative to CPTh  was suggested  the Chiral Confining theory (CCTh) \cite{18,19,20,21,22,23}
which contains both quark-antiquark and chiral meson degrees of freedom and provides the same
chiral relations e.g. the Gell-Mann, Oakes, Renner (GMOR) relations \cite{2}. What is more important the CCTh allows
to calculate the chiral meson masses, chiral condensate and the decay constants via the string tension and the quark masses, and take into account the external electromagnetic fields in agreement with numerical lattice data \cite{3*}.
It is interesting that the NJL approach which also takes into account the quark-antiquark d.o.f. provides results in better agreement with lattice data \cite{23*,23**} than the standard CPTh method.
Below we exploit the relativistic QCD Green's function and Hamiltonian formalism to calculate the chiral condensate in the external magnetic field and at nonzero temperature.

Here one needs the QCD Hamiltonian, incorporating confinement with a possible inclusion of external magnetic field. This QCD Hamiltonian and the meson wave functions were derived in the well developed Fock--Feynman-- Schwinger
formalism \cite{24,25,26}, which gives the masses and the wave functions of the lowest and high excited mesons with a good accuracy \cite{27}, as well as the Green's functions, necessary for the chiral condensate. To define the chiral condensate  the essential points are the basic relativistic theory with field correlators \cite{28} and the scalar character of the confinement interaction, which was proved in \cite{29}.

The main purpose of this paper is the study of the chiral condensate $\Sigma$ at nonzero temperature and magnetic field, deriving the basic expression for $\Sigma(T,e_q B)$, and comparing it with the lattice data.  In the previous paper \cite{30}, based on the CCTh, it was argued that the chiral condensate is proportional to the string tension in some power and, hence, both  vanish at the same temperature $T_c$. This fact can be approximately seen in the lattice results \cite{31}. In this paper we will show that for light quarks the chiral condensate is proportional to the
string tension to the power $3/2$ and this result is also supported by the lattice data, see \cite{11}.

The extension of previous results to the nonzero constant magnetic field will be done here, using the technic developed by our group in \cite{10*,32,33,34,35}. This approach  has allowed to compute behavior of the light quark condensate in the magnetic field below $eB=1$~GeV$^2$. Here we also generalize that analysis for arbitrary magnetic field and changing temperature $T$ so that the quark  condensate is becoming the function of both $T$ and magnetic field $eB$ for all quarks $u,d,s$.
For these quarks there are now precise lattice calculations of the quark condensates in wide
 regions of $T, e_q B$ and it allows to compare our results with these data and define the accuracy of our approach. The previous analysis of the $eB$ dependence of the chiral characteristics at zero temperature  \cite{10*,32,33,34,35}
has shown a good agreement of our theory with lattice data and strong disagreement with the standard chiral perturbation theory (CPTh), so that it becomes important to check this question again in a wider scope of both values,  $T$ and $eB$, which will be changed here in  wide intervals. At the same time this can be an additional proof of our dynamical QCD theory with confinement, created by the field correlators
\cite{28}, and the chiral symmetry breaking, directly connected with the scalar character of the resulting confinement. As it was shown in the recent paper \cite{30},
the phenomena of confinement and CSB, combined in the effective Lagrangian, are indeed closely connected and can be written in the form of CCTh \cite{18,19,20,21,22,23}.
In this paper we will make one step further and express the chiral condensate effectively (after renormalization) via the modulus square of the ground state $q\bar q$ singlet wave function,
$\Sigma={\rm~ const} |\phi_{(n=0)}(0)|^2$, also for nonzero $T,eB$. Remarkably, as will be shown here, this relation is supported by numerous lattice data up to largest magnetic fields.

In the CCTh physical picture the ``fundamental'' d.o.f. belong to an individual quark that moves in the confining vacuum. Then the gauge invariant definition of the Dirac operator eigenmode requires construction of a white object-- that is the Green's function of the quark, propagating along a closed trajectory in the 4$D$ Euclidean space-time.
The associated Wilson loop obeys the area law, which produces non-perturbative quark effective mass, expressed via the string tension, which is nonzero even for zero current quark mass. This effect dynamically breaks the chiral symmetry and generates the quark condensate.
The accurate calculation of the quark and the meson Green's functions is performed  with the Fock--Feynman--Schwinger Method \cite{24,25,26,27}.

The plan of the paper is as follows. In section 2 we derive the basic expression of the chiral condensate via the sum over $q\bar q $ states and finally via the ground state wave function. In
section 3  the explicit form of the wave functions and $q\bar q$ eigenvalues is given in terms of the string tension, the quark masses, and external magnetic field. In section 4 the numerical estimates of the chiral condensate as a function of the quark mass $m_q$, external magnetic field $eB$, and the temperature $T$ are given and compared with lattice data. The last section is devoted to conclusions and an outlook.

\section{From scalar confinement to spontaneous violation of chiral symmetry} \label{SectConf}

In \cite{18} we have shown that the vacuum averaging in the confining vacuum fields (in the lowest cumulant order) leads to the set of two connected equations, defined  for the averaged quark Green's function and the vacuum averaged effective quark mass operator, and this expression will be used below to calculate the quark condensate. For that we apply the cumulant expansion directly to QCD Lagrangian,
$$
    \int\cD A \exp\int d^4x \left(\psi^\dag\slashed{A}\psi + \frac12\tr F_{\mu\nu}^2\right) =
    \braket{\exp\int d^4x \left(\psi^\dag\slashed{A}\psi\right)}_A = $$
 $$   = \exp\left(\int d^4x \psi^\dag(x)\gamma_\mu\bbraket{A_\mu(x)}\psi(x)+\right. $$\be
                \left. + \frac12\int d^4xd^4y \psi_a^\dag(x)\gamma_\mu\psi_b(x) \psi_c^\dag(y)\gamma_\mu\psi_d(y)
                \bbraket{A_\mu^{ab}(x)A_\mu^{cd}(y)}+\ldots
           \right),
\label{1}
\ee
where the Latin indices refer to the fundamental gauge group. As the next step, using the Fock--Schwinger gauge,
\be
    A_\mu(x) = \int_0^1du~ux_\nu F_{\nu\mu}(ux)
\label{2} \ee
we get
\be
    \bbraket{A_\mu^{ab}(x)A_\nu^{cd}(y)} = \frac{\delta_{bc}\delta_{ad}}{N_c} \int_0^1du~ux_\lambda \int_0^1dv~vy_\rho
    \bbraket{\tr F_{\lambda\mu}(ux)F_{\rho\nu}(vy)}.
\label{3} \ee
The latter correlator may be expressed via the standard correlators $D(z),D_1(z)$ \cite{28} and here for simplicity we keep only the confining component $D(z)$,
since the non-confining interaction does not generate quark condensates:
\be
    \braket{\tr F_{i4}(x)F_{j4}(y)} = -\delta_{ij}D(x-y).
\label{4}\ee

As a result, one obtains the effective quark action in the confining vacuum,
\begin{align}
    \mathcal{L}_\text{eff}(\psi^\dag, \psi) = &\int d^4x\psi^\dag(x)(-i\slashed{\partial}-im_q)\psi(x) +\\\nonumber
    &+ \frac{1}{2N_c}\int d^4xd^4y~\psi^\dag_a(x)\gamma_4\psi_b(x)~\psi_b^\dag(y)\gamma_4\psi_a(y)~J(x,y),
\end{align}
\label{5}
where
\be
    J(x,y) = \int_0^1du~x_i\int_0^1dv~y^i D(ux-vy).
\label{6}\ee

For this effective action the quark propagator $(S(x,z))$ and the effective quark mass $M_q(x,y)$ satisfy the set of equations,
\begin{gather}
    iM(x,z) = J(x,z)\gamma_4 S(x,z)\gamma_4,\\\nonumber
    (\slashed{\partial}+m)S(x,y) + \int d^4z M_q(x,z)S(z,y) = \delta(x-y).
\label{7}
\end{gather}
In what follows we will use the local approximation of the effective mass operator,
\be
    M_q(x,z) \approx ~M_q(x)\delta(x-z),
\label{8}\ee
that yields the approximation for the effective quark propagator,
\be
    S^{-1} = \slashed{\partial}+m_q+M_q(x),
\label{9}\ee
To proceed we follow the reasoning of \cite{21} and write the chiral condensate as
 \be\Sigma_q =  N_c |tr S_{xx}|;~~ S_{xy}=\left( \frac{1}{m_q+ \hat M_q + \slashed \partial}\right)_{xy}.
\label{10}\ee
Now one can associate $S(x,x)$ with a quasi-circular loop quark trajectory with $M_q(r)$ being the confining interaction between quarks at the opposite sides of the loop
(and hence between quark and antiquark). If one neglects the gluon-exchange interaction, then the confining potential provides $\lan M_q(r)\ran $ and the average $\lan r\ran $ with accuracy 10\%,
namely, $\lan M_q(r)\ran = \sigma \lan r\ran$ (with $\lan r\ran$ is the average size of the loop) and for a light quark with $m_q=0$  the size $\lan r\ran = 1.578\sigma^{-1/2}$.

Dividing  and  multiplying (\ref{11}) by $(m_q+ M_q(x) - \slashed \partial)$, one obtains
$$
\Sigma_q= tr<(m_q+ M_q+ \slashed\partial)^{-1}(m_q+ M_q- \slashed\partial)(m_q+M_q-\slashed\partial)^{-1}=$$\be = N_c \lan m_q+M_q\ran \int d^4yG(x,y)= N_c\lan m_q+M_q\ran f(\sigma,m_q),
\label{11}
\ee
 and using finally the expansion of the PS $q\bar
 q$ Green's function $G(x,y)$ in the series of eigenfunctions \cite{19,20,21,22} one obtains
 \be
 \Sigma_q= N_c \lan m_q+ M_q\ran  \sum_n \frac{|\phi_n(0)|^2}{M_n},
 \label{12}\ee
 where $\phi_n(r)$ and $M_n$ are the pseudoscalar (PS) $q\bar q$ eigenfunctions and eigenvalues of the QCD Hamiltonian.

The QCD Hamiltonian with confinement, perturbative and spin-dependent interaction, its solutions and the
Green's functions were studied in \cite{24,25,26} and here we give the expressions for the function $\phi_n$ and the eigenvalues $M_n$ from \ref{12}.  Although true singlet mass $\bar M_n$ includes the contributions of confinement $M_n(\sigma)$, the hyperfine $\Delta_{HF}$ and the self-energy interaction, $\Delta_{SE}= -\frac{4\sigma}{\pi\omega_n}$, here we keep for simplicity only $M_n(\sigma)$ and  write for the quantities, entering in \ref{12},

\be
M_n(\sigma)= \frac{m_q^2}{\omega} + \omega + \epsilon(\omega),
\epsilon_n(\omega)= \omega^{-1/3} \sigma^{2/3} a(n).
\label{13}\ee
Here $\omega_n$ is the average quark energy $\lan\sqrt{\vep^2+m_q^2}\ran$, calculated from the condition $\frac{\partial M_n}{\partial \omega}= 0$. For $m_q=0$ one has $\omega_n= \sqrt{\sigma}(a(n)/3)^{3/4}$.
The numerical values of $a_n,\omega_n$ are  known and  given in \cite{27}, e.g. $a_0= 2.338$.
It is important that for light quarks $\bar M_0 \approx 1/4 M_0$.
For $m_q=0$ and the confining, perturbative, and spin-spin interactions one obtains the following  values of  $\bar M_0,\omega_0$ for the ground state:

\be
M_0= 0.40~{\rm GeV}, ~\omega_0= 0.35~{\rm GeV}.
\label{14}\ee
In a similar way the values of the wave functions at the origin are found to be \cite{21,22},
\be
|\phi_n(0)|^2= \frac{\omega_n \sigma}{4\pi}
\label{15}\ee
More accurate values of $\phi_n(0), M_n$ can be found in \cite{27}.

However one can see that the series is not converging and should be renormalized. This renormalization process was described in \cite{30}, see Appendix 1 below , where it was shown that one can extract the divergent small distance contribution of the sum $f(\sigma,m_q)= \sum_n \frac{|\phi_n(0)|^2}{M_n}$ which yields $f\sim M_0^2$.
Now for small $m_q$ one can connect ${\bar M_0}^2 =~{\rm const} ~\sigma = {\rm const}'~\frac{|\phi_0(0)|^2}{\bar M_0}$.
Finally, approximating $\lan M_q(r)\ran  ={\rm  const}~\bar M_0 $, one obtains the equation which will be exploited below for small $m_q \ll \sqrt{\sigma}$ and arbitrary $T,eB $
\be
\Sigma(m_q,T,eB)={\rm  const}~|\phi_0(0)|^2.
\label{16}\ee
Then using (\ref{15}), one obtains
\be\Sigma_q= c_0 |\phi_0(0)|^2= c_0 \frac{|R_0(0)|^2}{4\pi}
\label{17}\ee
where $R_0(r)$ is the ground state radial singlet $q\bar q$ wave function.
Of the special interest is the case of the quark condensate in the external magnetic field $\Delta_q(eB)$, considered in the present formalism in \cite{32,33,34,35}, when
one can also  use the form \ref{15}, however, in this case the wave function at origin has a different form (see \cite{32}), since the $q-\bar q$ wave
functions and eigenvalues depend on the quark-antiquark spin projections $\nu_1,\nu_2$ on the magnetic field. Then, as shown in \cite{32,33,34},

$$
|\phi^{(+-)}_{n_perp=0,n_3}(0)|^2 = \frac{\sqrt{\sigma(e_q^2B^2+\sigma)}}{(2\pi)^{3/2}},$$\be
|\phi^{(-+)}_{N_oerp=0,n_3}(0)|^2 = (\sigma^2 c_{-+})^{3/4} \sqrt{1+ \left(\frac{e_q B}{\sigma}\right)^2 1/c_{-+}}.
\label{18},\ee
where
\be
 c_{-+}(B)=\left(1+\frac{8e_q B}{\sigma}\right)^{2/3}.
\label{19}\ee
Thus now we have all necessary equations to estimate quark condensate, both in zero and nonzero
external magnetic field. Also the case of the nonzero temperature $T$ can be considered, if we take into account that the string tension depends on $T$ -- it will be done in  next section.

\section{Numerical estimates and comparison to lattice data}

We start with the simplest example of the quark condensate for the zero mass quarks which can be
associated with $\Sigma_u,\Sigma_d$. Using (\ref{17}) and  \cite{27} for the light quark theoretical values  of $R_{1s}^l(0)= 0.376(8) $ GeV $^{3/2}$, one obtains
\be
\Sigma_u= \Sigma_d= c_0 (227~{\rm  MeV})^3,
\label{20}\ee
which can be compared to numerous lattice data \cite{4,5,6,7,8,9,10,11,12,13,14,15,16,17}, e.g. in \cite{30} $\Sigma_l= (283(2)$ MeV$)^3$. This implies that $c_0= 1.93$. In a similar way
we estimate the strange quark condensate, taking in (\ref{27}) $R_{1s}^s= 0.432$ GeV$^{3/2}$, and find
\be
\Sigma_s= c_0  \frac{|R_{1s}^s(0)|^2}{4\pi}= (309~{\rm  MeV})^3 .
\label{21}\ee
 This should be compared with the lattice result
from \cite{15} -- $\Sigma_s= (290(15)~{\rm MeV})^3$ demonstrating a good agreement of our equation (\ref{17})
with lattice data.

We now turn to the case of the quark condensate $\Sigma$ in the magnetic field, comparing our
theoretical prediction within our method \cite{32,33,34,35}, see Appendix 2 for details
\be
\Sigma_l(eB)= N_c\lan m_l+ M_0\ran\sum_n \left(\frac{|\phi^{(+-)}_n (0)|^2}{2 m^{(+-)}_n} + \frac{|\phi^{(-+)}_n (0)|^2}{2 m^{(-+)}_n}\right).
\label{22}\ee
We now make the same series of approximations as in (\ref{16}),(\ref{17}), replacing $\lan m_l+ M_0\ran$ by $c_0 m^{(+-)}$ and keeping in the sum in (\ref{22}) only the leading first term with $n=0$ and obtain
\be
K_q(T=0,e_q B)= \frac{\Sigma_l(eB)}{\Sigma_l(0)}= \frac{1}{2}\left[\sqrt{1+(e_q B/\sigma)^2}+ \sqrt{1+\frac{(e_q B/\sigma)^2}{c_{-+}}}\right],
\label{23}\ee
where $c_{-+}= (1+8\frac{e_q B}{\sigma})^{2/3}.$
Numerical results of $K_u(0,e_u B), K_d(0,e_d B)$ are given below in Tables 1 and 2.

\begin{table}[!htb]
\caption{Quark condensate for u quark in the magnetic field at zero temperature from (\ref{23}) in comparison with lattice data of \cite{11}}
\begin{center}
\label{tab.01}
\begin{tabular}{|l|c|c|c|c|c|c|}
\hline
$eB$(in GeV) &0 &0.2 &0.4 &0.6 &0.8 &1.0 \\
$K_u(0,e_u B)$(\ref{23}) &1 &1.16 &1.485 &1.86 &2.28 &2.69 \\
$K_u$ from \cite{11} &1 &1.19 &1.51 &1.86 &2.23 &2.59 \\
\hline
\end{tabular}\end{center}\end{table}
in a similar way we obtain the results for the d quark, shown below in the Table 2.

\begin{table}[!htb]
\caption{The same as in the Table 1 but for the d quark condensate}
\begin{center}
\label{tab.02}
\begin{tabular}{|l|c|c|c|c|c|c|}
\hline
$eB$ (in GeV) &0&0.2&0.4&0.6&0.8&1.0\\
$K_d(0,e_d B)$ (\ref{23}) &1 &1.096&1.15&1.307&1.48&1.67\\
$K_d$ from \cite{11} &1 &1.1 &1.23 &1.4 &1.57 &1.73\\
\hline
\end{tabular}\end{center}\end{table}

One can see a good agreement between theoretical and lattice data at the accuracy level of less than 10 percent.
One can also compare the average light quark condensate ratio $K^{+}=1/2(K_u(e_u B)+ K_d(e_d B))$
with the lattice data of \cite{11},which is done in Fig. 2 of \cite{35} together with the predictions of the
standard chiral theory. One can see  a remarkable agreement of our and lattice data of \cite{11}
and a strong disagreement between lattice data and results of standard chiral theory, which again
demonstrates the necessity of quark degrees of freedom in the chiral theory,  shown in \cite{10*}.  The resulting curves of $K^{-}$ in Fig. 3  of \cite{35} also
demonstrates a good agreement with the lattice data of \cite{11}.
To finalize our discussion of the $K_q(T=0,eB)$ behavior let us compare it with the large $eB$ lattice data of \cite{16}. For the largest magnetic field $eB= 3.4 GeV^2$ one has $K_u(T=0,eB= 3.4)= 7.7$ while the lattice data of \cite{16} yield $K_u(lat)= 7.5$. We can conclude that the zero temperature data are well described by our formalism of (\ref{23}).
At this point it is important to study the general trend of the large $eB$ chiral condensate,
which according to  \cite{14} might show a possible sign of the of the vacuum transition
connected to chiral magnetic effect. To this end we compare  the ratio $R_{lat}(eB)= \frac{\Delta\Sigma_u(eB)+ \Delta\Sigma_d(eB)}{\Delta\Sigma_u(1~{\rm GeV}^2)+ \Delta\Sigma_d(1~{\rm GeV}^2)} $
 from the lattice measurements in \cite{16,14} with our calculations according to (\ref{23})
 where $R_{th}(eB)= \frac{K_u(eB)+ K_d(eB)- 2}{K_u(1~{\rm GeV}^2)+K_d(1~{\rm GeV}^2)- 2}$.
 In this way we obtain $R_{th}(4~$ GeV$^2)= 4.39$ and $R_{th}(9~$ GeV$^2)= 11.56$, while the lattice data from \cite{14}) yield respectively (approximately from Fig 3 of \cite{14})
  $R_{lat}(4~$GeV$^2)= 4.8; R_{lat}(9$~GeV$^2)= 12.6$.
 One can see again a reasonable agreement, implying that our formalism, without any phase transitions,
 accurately describes the behavior at large $eB$.

Another interesting comparison with data of \cite{11} can be done for nonzero temperature T and
growing magnetic field $eB$. To this end one must exploit the behavior of the string tension $\sigma(T)$ with temperature, which can be found from the same data of \cite{11}.
Our basic quantity to be discussed below is the quark condensate at nonzero $T,eB$ normalized with
zero values  $T=0,eB=0$. We start with the nonzero $T,eB$ expression for $\Sigma$
\be
\Sigma(T,eB)= N_c c_0 (|\phi^{(+-)}_0(0)|^2 + \frac{m^{(+-)}}{m^{(-+)}}|\phi^{(-+)}_0(0)|^2)
\label{24},\ee
Exploiting the forms of $m^{(+-)},m^{(-+)}$ from \cite{32,33}, one arrives at the following expression for
the normalized quark condensate of light quarks in the magnetic field at nonzero temperature $T$

\be
K_q(T,e_q B)= \frac{\Sigma(T,e_q B)}{\Sigma(0,0)}= \left(\frac{\sigma(T)}{\sigma(0)}\right)^{3/2} A(T,e_q B)
\label{25},\ee
where
\be
A(T,e_q B)= 1/2 \left(\sqrt{1+\left(\frac{e_q B}{\sigma(T)}\right)^2}+\sqrt{1+\left(\frac{e_q B}{\sigma(T)}\right)^2
\left(1+\frac{8e_q B}{\sigma(T)}\right)^{(-2/3)}}\right).
\label{26}
\ee
To find the ratio $\frac{\sigma(T)}{\sigma(0)}$ one can use the lattice data from the Table of \cite{11} for the ratio $K_q(T,0)$ which does not depend on $e_q$ and is equal to $\left(\frac{\sigma(T)}{\sigma(0)}\right)^{3/2}$. The resulting numbers are given in Table 3.

\begin{table}[!htb]
\caption{The temperature dependence of the string tension $\sigma(T)$}
\begin{center}
\label{tab.03}
\begin{tabular}{|l|c|c|c|c|c|c|c|c|c|c|}
\hline
$T$(in MeV) & 0 &113&122&130&142&148&153&163&176&189\\
$K_q(T,0)$  & 1 &0.90&0.84&0.80&0.68&0.57&0.49&0.26&
0.08&0\\
$\sigma(T)/\sigma(0)$ & 1 &0.932&0.89&0.86&0.773&
0.687&0.62&0.407&0.185&0\\
\hline
\end{tabular}
\end{center}
\end{table}

We are now in the position to find the quark condensate at arbitrary $T,e_q B$ and we shall do it
for the fixed temperature $T= 113$ MeV and $eB$ in the range $(0,1)$ GeV$^2$, as shown below in the Table 4.

\begin{table}[!htb]
\caption{The condensate of the u quark in the magnetic field at the temperature $T= 113$ MeV}
\begin{center}
\label{tab.04}
\begin{tabular}{|l|c|c|c|c|c|c|}
\hline
$eB$(in GeV$^2)$ &0 &0.2 &0.4 &0.6 &0.8 &1.0\\
$K_{1/2}$ (\ref{25}) &0.9 &1.06 &1.39 &1.76 &2.17 &2.57 \\
$K_{1/2}$ \cite{11} &0.9 &1.05 &1.33 &1.68 &2.0 &2.35 \\
\hline
\end{tabular}
\end{center}
\end{table}
On can see a reasonable agreement on the level of 10 percent or better. To understand better the situation with the $T$-dependence of the $\Sigma(T,eB)$ one can compare the lattice data of \cite{17} $\Delta\Sigma_q(B,T)$ with our predictions of $K_q(T,e_q B)$ via the relation
\be
\xi(B,T)= \frac{\Delta\Sigma(B,T)}{\Delta\Sigma(B,0)}= \frac{(K_q(T,e_q B)- K_q(T,0))}{K_q(0,e_q B)-1}.
\label{27}\ee
One can see also  a reasonable agreement of our ratio $\xi(B,T)$ with the data of $\Delta\Sigma(B,T)$ from Fig. 3 of \cite{17}.
In the same way we can compare our basic equation for $K_q$ \ref{25} with the lattice data of \cite{13}.

Finally we can analyze  reasonable bounds of our approximations resulted in the our final (\ref{25}), where we adopted the point of view that the string tension depends only on the temperature $T$, $\sigma(T)$ even in the situation when the magnetic field $eB$ is nonzero. It is clear that the
critical temperature depends on magnetic field, $T_c= T_c(eB)$, and therefore for large magnetic fields and $T$ around $T_c(eB)$ the string tension should  also depend on the value of the magnetic
field. To check this assumption we now calculate the ratio $K_u(T=130$ MeV, $ e_u B)$ as defined in  (\ref{25}) and compare it with the accurate lattice data \cite{11}. The results are shown below in Table 5.

\begin{table}[!htb]
\caption{The magnetic field dependence of the chiral condensate for the temperature $T= 130$ MeV}
\begin{center}
\label{tab.05}
\begin{tabular}{|l|c|c|c|c|c|c|}
\hline
$eB$(in GeV$^2$) &0 &0.2 &0.4 &0.6 &0.8 &1.0 \\
$K_u $(\ref{25})  &0.8 &0.96 & 1.28 & 1.6& 2.03 &2.4\\
$K_u(lat)$, (\ref{11}) &0.8 &0.97 & 1.2& 1.41 &1.6 & 1.76\\
\hline
\end{tabular}
\end{center}
\end{table}

One can see in Table 5 a good agreement (within \%15) of theoretical and lattice data up to a value
of the magnetic field $eB= 0.8$~GeV$^2$ at our fixed temperature $T=130$~MeV, which implies that
we are approaching the critical region of the deconfining transition at the given temperature and magnetic field,$T_c(eB=0.8$ GeV$^2)$, where string tension depends on both $T,eB$.
Summarizing, we conclude that our method of  the chiral condensate calculation shows a good agreement with lattice
data in the confining region of the  $(T,eB)$ plane except for the narrow region near $T_c(eB)$.

\section{Conclusions}

The equations (\ref{23}),({\ref{25}),(\ref{26}) contain our basic predictions for the chiral condensate of light quarks
for arbitrary temperatures and magnetic fields. Also in the Tables 1,2,4,5 we have demonstrated good agreement of our results
with lattice data -- within ten percent accuracy, for all temperatures  with exception of
narrow region around the critical point $T_c(e_q B)$. We have also performed the extension to nonzero quark masses, given in (\ref{15})--(\ref{17}) in the case of zero magnetic field. The most general case  for nonzero quark masses, temperatures and magnetic fields is planned to consider for the future, as well as a more detailed comparison of theory and numerical data. Our result that the chiral condensate is proportional to the probability of the singlet quark-antiquark system to be at one point, seems quite natural, if one considers a simple picture of the chiral condensate, given by the small loop quark propagator, where the radius of the loop is decreasing for growing magnetic field and therefore the quark density at the center grows. In the same way for growing temperature and decreasing
$\sigma(T)$ (confinement) the loop radius is increasing and therefore at the center of the loop the density is decreasing. We have shown that the magnetic field adds to the linear confinement potential a parabolic attraction along the field axis,  and it is acting coherently with confinement.
At this point one can wonder about the possible singular behavior of chiral characteristics
in strong magnetic field in the framework of the chiral magnetic catalysis \cite{17*}. This topic is connected to the confinement behavior in the strong magnetic field studied in \cite{36} and is associated with the sea quark production in the strong magnetic field. The latter is also
 connected to the $O(\alpha_s)$ contributions in the meson masses and was discussed in detail
in the review paper \cite{37}, and as was shown in \cite{38} the perturbative mass correction stays finite and all effects including spin-spin interactions in mesons are not divergent at large $eB$ \cite{39}.
  Therefore one can expect a rather smooth behavior with increasing magnetic field, as confirmed by our equations (\ref{23}) in agreement with the lattice data.

Another point is the basic role, played by the scalar confinement in creating this nonzero probability in the quark loop discussed above, which is supported by data. It is also important that vector confinement cannot exist in our universe, see \cite{29}. It is interesting that the final expression for the chiral condensate via $|\phi(0)|^2$ (\ref{16}) strongly decreases when the scalar confinement
vanishes at $T= T_c$ and $|\phi(0)|^2$ is defined by the perturbative gluon exchange interaction
 proportional to $(m_q \alpha_s)^3$. In this way
our formalism is in agreement with the observed fact that chiral condensate and string tension
disappear at roughly the same temperature. Summarizing one say that the chiral condensate  is produced by confinement, but this is only the first part of the story. The next step is to obtain pion and other PS masses,as well as PS decay constants again from confinement. This is done within our Chiral Confining Lagrangian \cite{23}, producing among other things the GMOR relations.

The author is grateful to A.M.Badalian for useful discussions and comments, and to N.~P.~Igumnova
for important help in preparing the manuscript.

\vspace{2cm}

{\bf Appendix A1}\\

 {\bf  Renormalization of the sum over $q\bar q$ states}\\

 \setcounter{equation}{0} \def\theequation{A\arabic{equation}}

The closed propagator \(f(\sigma,m_q)\) is divergent because of the zero distance pole.
A meaningful result for the chiral condensate can be obtained by subtraction of the regularized pole value of the free propagator.
To do that, let us consider the propagator at a small (temporal) distance \(t_0\)
\be
    f(\sigma, 0, t_0) = \int_{t_0}^{+\infty}dt\sum_n e^{-M_nt}|\varphi_n^2(0)|.
\label{A1}\ee
With color-Coulomb and spin-spin interactions neglected, we utilize the string spectrum
    \(M_n^2=4\pi\sigma(n+3/4)\), \(|\varphi_n^2(0)| = \frac{\xi\sigma M_n}{4\pi}\) as a model for the spectrum and approximate the \(n\) summation with the \(dM\) integration
\be
    f(\sigma, 0, t_0) \approx \frac{\xi}{8\pi^2}\int_{M_0}^{+\infty}M^2dM\int_{t_0}^{+\infty}e^{-Mt}dt = \frac{\xi}{8\pi^2}(t_0^{-2}-M_0^2) +O(t_0).
\label{A2}\ee
Comparison with the small-distance free particle propagator behavior \(G(t)\sim(4\pi^2t^2)^{-1}\)
    fixes the parameter \(\frac{\xi}{8\pi^2}=\frac{1}{4\pi^2}\).
Subtraction of the ``temporal divergence'' leads us to a numerical result,
    which happens to be fairly close to the lattice data
\be
    f(\sigma,0) = -\frac{M_0^2}{4\pi^2} = -\frac{3\sigma}{4\pi},
\label{A3}\ee
and as a result
\be
    \Sigma_l = \frac{3N_c\sigma^2\lambda}{4\pi} = (268 \text{ MeV})^3 .
\label{A4}\ee

\end{document}